%% file: main.tex
\documentclass[conference]{IEEEtran}
\IEEEoverridecommandlockouts
\usepackage{cite}
\usepackage{amsmath, amssymb, amsfonts}
\usepackage{algorithmic}
\usepackage{graphicx}
\usepackage{textcomp}
\usepackage{xcolor}
\usepackage{supertabular}
\usepackage{subcaption}

\usepackage{orcidlink}

\begin{document}
    \title{Risk-based approach to the Optimal Transmission Switching problem
    \thanks{This work is part of the CRESYM project OptGrid, funded by RTE - Réseau de Transport d'Électricité.}
    }

    \author{\IEEEauthorblockN{Benoît Jeanson  } \IEEEauthorblockA{\textit{CRESYM}\\ Brussels, Belgium 
    \\ https://orcid.org/0009-0005-4352-2604
    }
    \and \IEEEauthorblockN{Simon H. Tindemans } \IEEEauthorblockA{\textit{Dept. of Electrical Sustainable Energy} \\ \textit{Delft University of Technology}\\ Delft, The Netherlands \\
    https://orcid.org/0000-0001-8369-7568}
    }

    \maketitle

    \begin{abstract}
        This paper deals with the secure Optimal Transmission Switching (OTS) problem in situations where the TSO is forced to accept the risk that some contingencies may result in the de-energization of parts of the grid to avoid the violation of operational limits. This operational policy, which mainly applies to subtransmission systems, is first discussed. Then, a model of that policy is proposed that complements the classical MILP model of the N-1 secure OTS problem. It comprises a connectivity and notably a partial grid loss analysis for branch outage contingencies. Finally, its application to the IEEE 14-bus system is presented. Solutions similar to those observed in operation are reached by the algorithm, notably revealing the preventive-openings-cascade phenomenon. 
    \end{abstract}


    \input{terminology.tex}
    \input{introduction.tex}
    \input{methodology.tex}
    \input{modeling.tex}
    \input{results.tex}
    \input{conclusion.tex}
    \section*{Acknowledgment}
    The authors express their thanks to 
    RTE Research and Department team,  especially Nourredine
    Henka, Sami Tazi, Patrick Panciatici and Lucas Saludjian for helpful discussions.

    \bibliographystyle{IEEEtran}
    \bibliography{IEEEabrv,references_bibtex}
\end{document}

%% file: terminology.tex
\section*{Nomenclature}
\label{ss:Formulation}

\noindent
We consider a graph $\mathcal{G}$ and denote vectors in lowercase bold  $\pmb{u}$, and matrices  in uppercase bold $\pmb{A}$. 

\subsection*{Functions and operators}

\noindent
\begin{tabular}{p{1.5cm}p{6.5cm}}
    $\odot$            & Hadamard product: elementwise multiplication \\
    $\langle .,.\rangle$ & Inner product                               \\
    $\delta_{c,i}$     & Kronecker function: 1 if $c=i$; 0 otherwise       \\
    $\text{org}(e)$    & Origin vertex index of edge $e$              \\
    $\text{dst}(e)$    & Destination vertex index of edge $e$         \\
    $\text{inc}(i)$    & Incident edges of vertex $i$\\
    $\text{opp}(e,i)$  & For an edge $e$ connected to a vertex $v$, the vertex on the other end of $e$, distinct from $v$.
\end{tabular}
For example, if $e \in \text{inc}(v)$, then $v$ is either in $\text{org}(e)$ or $\text{dst}(e)$. Moreover, if $v=\text{org}(e)$, then $v = \text{opp}(e, \text{dst}(e))$.
\subsection*{Sets}
\noindent
\begin{tabular}{p{1.5cm}p{6.5cm}}
    $\mathcal{V}$         & indices of the vertices or buses of $\mathcal{G}$    \\
    $\mathcal{E}$         & indices of the edges or branches of $\mathcal{G}$    \\
    $\mathcal{C}^{\star}$ & edge indices of contingencies, $\mathcal{C}^{*}\subset \mathcal{E}$.\\
        $\mathcal{C}$         & edge indices of contingencies including $0$ for none
\end{tabular}

\subsection*{Matrices, vectors and scalar}

\noindent
\begin{tabular}{p{1.5cm}p{6.5cm}}
    $\pmb{A}$           & The incidence matrix of $\mathcal{G}$. $\pmb{A}_{ve}=0$ if the edge $e$ does not connect the vertex $v$, $1$ if it is oriented toward $v$; $-1$ otherwise. \\
    $\pmb{0}, \pmb{1}$  & vectors filled respectively with 0 and 1                                                                                                             \\
    $\pmb{g},\ \pmb{d}$ & Generations, loads on buses                                                                                                                       \\
    $\pmb{p}$ & $\pmb{p} = \pmb{g} - \pmb{d}$ Net injections on buses                                                                                                                       \\
    $\pmb{b}$           & Susceptances of branches                                                                                                                             \\
    $\bar{\pmb{f}}$ & Flow limits on branches\\
    $\pmb{v}$           & Binary vector indicating branch openings in the base case. $v_{e}= 1$ if edge $e$ is open, 0 otherwise   
\end{tabular}

\noindent

\medskip
\noindent
For a contingency $c \in \mathcal{C}$:

\noindent
\begin{tabular}{p{1.5cm}p{6.5cm}}
    $\hat{\pmb{g}}_{c},\ \hat{\pmb{d}}_{c}$ & Post-contingency generations and loads                                                                     \\
    $\pmb{f}_{c}$                           & Directed flows in the edges                                                                                          \\
    $\pmb{\phi}_{c}$                        & Voltage phase angles in the buses                                                                           \\
    $\pmb{p}^{\star}_{c}$                   & Net injections in N-1 mirror graph                                                                          \\
    $\pmb{f}^{\star}_{c}$                   & Flows in N-1 mirror graph                                                                                   \\
    $\pmb{w}_{c}$                           & Binary vector indicating that the branch is opened either in the base case ($\pmb{v}$) or because of the contingency $c$                                                                     \\
    $\pmb{\mathfrak{p}}_c$      & Probability of occurrence of contingency $c$\\
    $\pmb{\pi}_{c}$                         & Binary vector indicating the energized status of buses. $\pmb{\pi}_{c,i}=1$ if bus $i$ is energized, 0 else \\
    $\psi_{c,i,e}$                          & Variable indicating whether the bus connected to bus $i$ via edge $e$ is energized\\
    $\sigma_{c}$                            & Generation scaling factor
\end{tabular}

\medskip

%% file: introduction.tex
\section{Introduction}

One of the primary concerns of Transmission System Operators (TSOs) is ensuring the security of the power system. To achieve this, they must anticipate potential events that can compromise it while simultaneously striving to maintain economic efficiency.
When it comes to security policies, the cornerstone is the N-1 rule. Sometimes, it is
reduced to the statement that the consequences of the tripping of any single element
of the grid shall have no impact on the users of the power system. This is
actually a narrow view. In fact, the rule derives from a risk-based principle. As
contingencies are likely to occur, what is at stake for the operator is that any resulting
state of the system remains under control. In other words, the consequences of
any likely tripping of a branch shall be mastered. Furthermore, to align the risks to the
same level, the more likely a contingency is, the less acceptable its
consequences in case of occurrence. So, the criteria the operator follows
are based on the definition of the limits between an acceptable and an unacceptable
consequence with respect to each contingency considered.

Keeping flows or voltages within security limits is one of the key
criteria. Losing a part of the grid and de-energizing users could also be
considered unacceptable, but is it realistic? In fact, enforcing the latter constraint
implies that the real-time operator must avoid \emph{at all costs} losing customers
as a result of a single tripping. This may involve operations on the grid itself,
for example, by implementing switching actions or activating redispatch or curtailment of customers, which rapidly becomes expensive. 

EU regulation \cite{eu-2017/1485} does not contain such
standards. \emph{Loss} is only explicitly mentioned in the description of the security
states where it is introduced
for the black-out state, which is defined by the \emph{loss} of more than 50\% of
the total consumption of the control area of the TSO. The fact that there is no specific
standard reflects at least that there is no requirement to not lose any part of the grid,
and opens to the interpretation that the regulation admits this may happen. Moreover, there may be operational scenarios where solutions that avoid all customer disconnections do not exist.

This paper considers situations where no solution is available, or where the costs associated with the mitigation of potential loss of parts of the grid are prohibitive. Such cases
are addressed in the context of the Optimal Transmission Switching (OTS) problem.

The OTS problem consists in finding the combination of branch openings in the grid
that optimizes a given objective function. Many optimization approaches to the
OTS problem have been developed. \cite{fisherOptimalTransmissionSwitching2008} by
Fisher~\emph{et al.}~is the first DC-OPF (Direct Current Optimal Power Flow) based on an MILP (Mixed-Integer
Linear Program) approach to it with the goal of solving \emph{congestion}.
This pioneering study lays the foundation for numerous subsequent efforts that employ
the DC approximation to solve the OTS. In this model, the switching problem is
formulated using a big-M approach and the necessary flexibility -- when no
switching scheme is enough to cope with the congestion -- is provided by the OPF
equations that aim at minimizing generation adjustment cost. Hedman~\emph{et al.}~extended
that approach in \cite{hedmanOptimalTransmissionSwitching2009} to include \emph{security
analysis} and incorporate the N-1 rule. The algorithm does not permit the loss of load or generation, but it does allow electrical islands either in the
base case or after a contingency if the resulting islands satisfy all the constraints.
 In practical operation, islanding should
only be considered in very particular power systems that are designed to be sustainable
after a grid split, which implies the fulfillment of a lot of
requirements in terms of balancing, control, and stability. To avoid islanding, various recent papers therefore  take into account a connectivity constraint, for which various approaches have been proposed (\cite{ostrowskiTransmissionSwitchingConnectivityEnsuring2014}, \cite{dingMixedIntegerLinearProgrammingBased2018}, \cite{hanEnsuringNetworkConnectedness2021}, \cite{liConnectivityConstrainedMILP2021}). 

The contribution of this article is threefold. First, a discussion of the N-1 rule
is presented and consequences of including it in operation are presented. Then
a deterministic optimization model that reflects that risk-based operational policy is developed. At its
core, the connectedness of the network in the base case and in the N-1
situations is assessed. Finally, its application to the IEEE 14-bus system is
analyzed.

%% file: methodology.tex
\section{A risk-based approach to the OTS}

\subsection{Grid designs and security}

The authors of \cite{eggletonNETWORKSTRUCTURESUBTRANSMISSION1993} expose various
designs of power grids and analyze the implications in terms of security level. Underlying
their rationale is the principle that the lower the consequences, the more
acceptable they are. By applying that logic, the distribution systems are
operated radially, though such an operating scheme puts all the connected users under
the risk of being shut down by the contingency of any element of the string of elements
connecting it to the source substation. In contrast, transmission systems dealing
with larger areas are exposed to bigger consequences and foster a meshed operation
to secure each substation. Therefore, the bulk transmission grid is operated fully
meshed, whereas underneath, subtransmission parts are operated looped or meshed,
some parts being operated as pocket under one single feeding point, or groups connected
to multiple feeding points.

These parts of the transmission grid are more prone to being operated with insecure
areas. This is especially the case where the grid is weakly developed, or if the
development could not follow the settlement of grid users, during maintenance
periods that require outages of nonredundant elements, or following a contingency.
One must accept that some areas of the system -- even if internally meshed --
would only be connected to the main part of the network through a single branch.
In that case, should this branch trip, the whole area is de-energized.

But that insecurity does not mean that the transmission system operator lacks control.
The consequences of each potential contingency must be clearly identified and
mastered. No tripping should result in violation of physical limits, cascading
effects, or grid collapses. Should a given contingency lead to loss of a part of
the system with loads and/or generations, the extent of the consequences must be
assessed \emph{a priori} and weighted against its probability of occurrence.

\subsection{Implication to the transmission switching design process}

This aspect of policy must be considered by operators when designing their
operational strategies and especially when designing the switching patterns. Whereas for some assets, the possibility of temporarily overloading a branch may pose acceptable risks, this may not be the case for other assets. We consider the latter case, where even if the probability is low, the consequence is considered unacceptable as this could destroy assets or worse still expose people around to potentially fatal outcomes.

Consider the case of a subtransmission area connected to the bulk grid through two branches. Should the tripping of either one lead to overloading of the other, the operator must take preventive actions. One option consists in involving grid users in the area and requesting a change in their injections, which is costly and would need to be done irrespective of whether the contingency occurs. The other consists in splitting the area into two disconnected pockets, each hanging on one branch. Now, if a tripping occurs on one of these branches, the corresponding pocket is lost, but there would be no forbidden overloading.
That is an application of the security/cost trade-off: to decide between spending money for redispatching and ensuring security or taking the risk of losing an area, which would be very unlikely, but with potential significantly higher costs. In the latter case, this may imply the search for an optimum among all the possibilities of splitting the area into two pockets. This is the focus of this paper.

\begin{figure}
    \includegraphics[width=\linewidth]{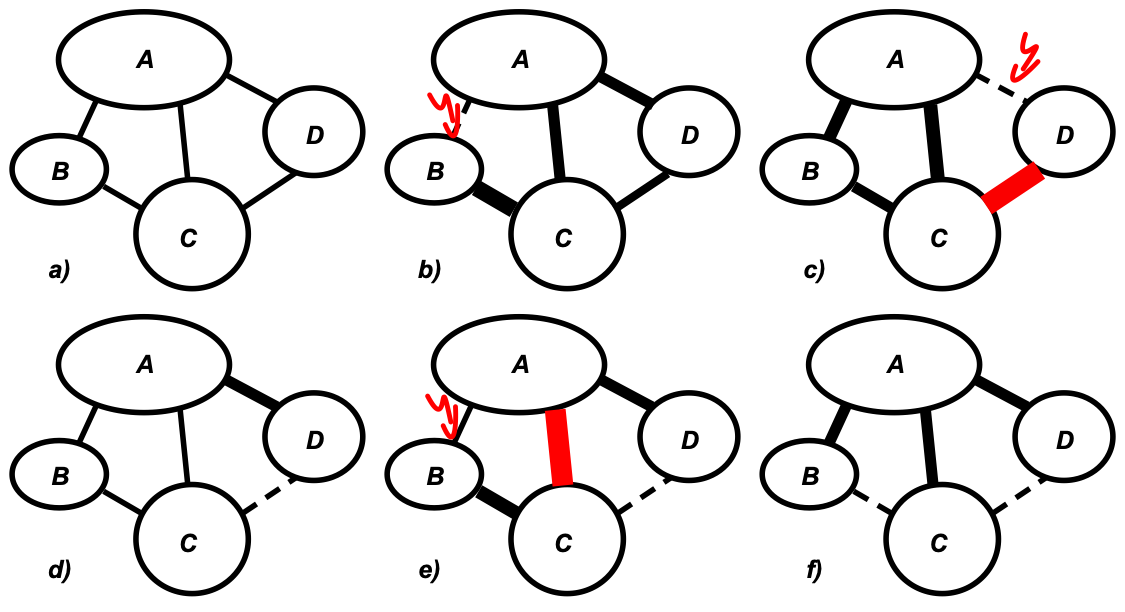}
    \caption{Preventive-openings-cascade. Each bubble represents a portion of a grid
    a) the initial pattern;
    b) after the tripping of $AB$, no overloading and $B$ is secured;
    c) the preventive opening of $CD$;
    d) the tripping of $AB$ now leads to an overloading of $AC$;
    e) the tripping of $AD$ results in an overloading on $CD$ which implies the opening of $BC$
    f) final situation avoiding post-contingency overloadings.}
    \label{fig:un-meshing}
\end{figure}

The next illustration pushes a similar rationale in a more complex context and
highlights the preventive-openings-cascade phenomenon. Figure \ref{fig:un-meshing} shows the chain
of implications of taking into account the consequence of contingencies. An opening that weakens its neighboring area requires another opening. Repeating this rationale results in a topology scheme with
separated pockets, each one hanging on a single branch.
It is worth noting that the grid may be meshed within pockets, so
that inner trippings do not necessarily result in de-energizations.

\subsection{Application to the OTS}
The OTS aims to find an optimal solution that aligns with the risk-based principle underlying the N-1 rule. It therefore is necessary to integrate the following conditions:
\begin{itemize}
    \item there must be no overloading in the base case;
    \item the risk associated with the tripping of a single branch is considered too high, if it leads to overloading, therefore, there must be no overloading after any tripping of a single branch;
    \item if no other option is available, or if the cost of mitigating exposure is prohibitively high, certain parts of the grid may face the risk of being de-energized as a result of the tripping of a single branch;
    \item the risk of being de-energized must be minimized by applying the switching schemes that limit as much as possible the exposure to that risk.
\end{itemize}

%% file: modeling.tex
\section{MILP modeling}
\subsection{Principles}
This section develops an MILP formulation of the approach described above. 
The aim is to determine preventive branch openings, if any, that will optimally reduce system risk. These are described by a binary vector $\pmb{v}$, indicating the pre-contingency status of each branch (1 is open). 

The base case is supposed to be balanced and the model shall ensure its connectedness
(\ref{ss:base case connectedness}). On the other hand, in N-1 the connectedness is
assessed to identify for each contingency which part of the grid, if any, is de-energized
(\ref{ss:N-1 Connectedness}). The generators remaining in the energized area are
adjusted to ensure the balance of the N-1 case (\ref{ss:balancing}).
Finally, a secured DC power flow (\ref{ss:dcpf}) is modeled to assess the flows in
the branches, and constraints are added to include their flow limitations. The cost
function aiming at minimizing the risk taken is described in \ref{ss:cost function}.

Some formulations are not directly applicable as MILP for they involve logical operations
or Hadamard product of variables. They are first exposed in their concise form
for simplicity and readability, and the steps used to translate them into MILP formulations are given in \ref{ss:bigM}.

\subsection{Base case connectedness}
\label{ss:base case connectedness} In order to ensure the connectedness of the grid
in the base case, (\ref{eq:base case connectedness}) implements a method inspired by \cite{dingMixedIntegerLinearProgrammingBased2018} involving
a fictitious mirror graph. A subscript $0$ is used to indicate the pre-contingency base case.
\begin{subequations}
    \allowdisplaybreaks
    \begin{align}
        \pmb{f}^{\star}_{0}\odot \pmb{v} & = 0 \label{subeq:connecteness branch opening}                     &                                                                                      \\
        \pmb{A}\pmb{f}^{\star}_{0}       & = \pmb{p}^{\star}_{0}\label{subeq:base connectedness conservation} \\
        \
 p^{\star}_{0,i}               & = 1                                                               & \quad \forall i \in \mathcal{V}\backslash \{s\} \label{subeq:base connectednes load} \\
        p^{\star}_{0,s}                  & = - (\left|\mathcal{V}\right| - 1)                                & \label{subeq:base connectedness source}
    \end{align}
    \label{eq:base case connectedness}
\end{subequations}
The mirror graph has the same vertices and edges as the graph of the grid. If one branch is open in the grid, the corresponding edge is open in
the mirror graph, and its flow is null (\ref{subeq:connecteness branch opening}).
In this mirror graph, the only law that applies is the principle of conservation (\ref{subeq:base
connectedness conservation}). One vertex with the index $s$ is chosen as the
source -- a virtual generator. All others consume $1$ (\ref{subeq:base
connectednes load}). Thus, if the graph is connected, there is a connected path between
the source vertex and any of the others, and by application of the principle of
conservation, the source vertex shall produce $\left |\mathcal{V}\right|-1$,
which is ensured by (\ref{subeq:base connectedness source}). Only a connected graph can satisfy these constraints.

\subsection{N-1 Connectedness handling}
\label{ss:N-1 Connectedness} 
We consider only branch opening contingencies; hence we use the convention that contingency $c\in\mathcal{C}^{*}$ corresponds to opening of branch $c$. A new binary vector $\pmb{w_c}$ is introduced that combines
the switching status of the branches for the contingency $c$. $w_{c,e}$ has value 1 if the edge $e$ is open either preventively by the OTS or because of the contingency; it is 0 otherwise:
\begin{equation}
    w_{c,e}= v_{e}
    \vee \delta_{c,e} \quad \quad  \forall \left(c,e\right) \in \mathcal{C}\times\mathcal{V}
\end{equation}
Note that for index $c=0$ (the base case), $\pmb{w}_{0}=\pmb{v}$.

In the risk-based approach, after a contingency, part of the grid may be
deenergized. As for the base case connectedness, a fictitious mirror graph is
implemented (\ref{eq:OTS N-1 connectedness}) for each considered contingency in
$\mathcal{C}$. However, the goal is no longer to ensure connectedness, but rather
to identify the buses that would be disconnected from the Main Connected Component
(MCC) of the grid after the trip. The MCC is defined as the set of nodes connected to the mirror grid source vertex. For simplicity, the same vertex $s$ chosen for the base case is used here, but more elaborate definitions could be used, including the use of a contingency-specific source vertex. 

$\forall c \in \mathcal{C}^{*}$:
\begin{subequations}
    \allowdisplaybreaks
    \begin{align}
        \pmb{f}_{c}^{\star}\odot \pmb{w}_{c}     & = \pmb{0}\label{subeq:N-1 connecteness branch opening}                           \\
        \pmb{A}\pmb{f}_{c}^{\star}               & = \pmb{p}^{\star}_{c}\label{subeq:N-1 connectedness conservation}                \\
        \pi_{c,i}                                & \in \{0, 1\}                                                                    & \forall i \in \mathcal{V}\label{subeq:N-1 connectedness pi declaration}  \\
        \pi_{c,s}                                & = 1 \label{subeq:N-1 connectedness set source}                                   \\
        p^{\star}_{c, s}                         & \in \{-(|\mathcal{V}| - 1) , ..., 0\} \label{subeq:N-1 connectedness source set} \\
        p^{\star}_{c,i}                          & = \pi_{c,i}                                                                     & \forall i \in \mathcal{V}\backslash \{s\} \label{subeq:N-1 pi consumers} \\
        \pi_{c,i}= \bigvee_{e \in \text{inc}(i)} & \left[ \pi_{c,\text{opp}(i,e)}\wedge(1-w_{c,e})  \right]                                      & \forall i \in \mathcal{V}\label{subeq:N-1 connectedness pi disjunction}
    \end{align}
    \label{eq:OTS N-1 connectedness}
\end{subequations}
Equations (\ref{subeq:N-1 connecteness branch opening}) and (\ref{subeq:N-1
connectedness conservation}) play the same role as (\ref{subeq:connecteness
branch opening}) and (\ref{subeq:base connectedness conservation}). To identify
the part of the grid that remains in the main connected component, an indicator variable
$\pi_{c,i}$ is introduced that is set to $1$ for the contingency $c$ if the vertex
$i$ is energized (\ref{subeq:N-1 connectedness pi declaration}), $0$ otherwise. Its value
for the source vertex is set to $1$ as it is per definition in the MCC (\ref{subeq:N-1
connectedness set source}). Now, a similar approach to that of the base case is applied
where the source vertex $s$ is the only one that generates connectedness flow (\ref{subeq:N-1
connectedness source set}), and all the other vertices consume it with a
value of $\pi_{c,i}$ (\ref{subeq:N-1 pi consumers}). So, when the vertex $i$ is
not connected to the MCC, as there is no connected path to the source vertex,
the vertex cannot consume and $\pi_{c,i}=0$. In fact, having $p^{\star}_{c,i}=1$
would contradict \ref{subeq:N-1 connectedness conservation} in the
de-energized area.
This only ensures that $\pi_{c,i}$ is set to 0 for de-energized buses. The next mechanism is
necessary to force it to be $1$ in the energized area. Starting from the source
vertex for which $\pi_{c,s}=1$, the status propagates by (\ref{subeq:N-1
connectedness pi disjunction}). 
Each bus $i$ must be energized if at least one of its connected ($w_{c,e}=0$) neighbors is also energized. 

\subsection{Balancing the energized area}\label{ss:balancing}
If a tripping leads to a de-energized area ($\pi_{c, i}=0$), then the generation and load in the remaining nodes must be
balanced to continue operation. 
We choose to balance by shifting all remaining energized generators
proportionally to their initial values:

$\quad \forall c \in \mathcal{C}^{*}$:
\begin{subequations}
    \begin{align}
        \pmb{\hat{d}}_{c}                                          & = \pmb{d}\odot \pmb{\pi}_{c}\label{subeq:balancing load}                 \\
        \pmb{\hat{g}}_{c}                                          & = \sigma_{c}\pmb{g}\odot \pmb{\pi}_{c}\label{subeq:balancing generation} \\
        \langle\pmb{1}, \hat{\pmb{g}}_{c}- \hat{\pmb{d}}_{c}\rangle & = 0 \label{subeq:balancing overall}
    \end{align}
    \label{eq:N-1 balancing}
\end{subequations}
$\pmb{\hat{g_c}}$ and $\pmb{\hat{d_c}}$ represent the adjusted values of the
generation and load after contingency. (\ref{subeq:balancing load}) sets the
load to 0 when the substation is de-energized and to its initial value otherwise.
Similarly, (\ref{subeq:balancing generation}) applies to generation, including the generation scaling factor $\sigma_{c}$ that ensures that the system is balanced \eqref{subeq:balancing overall}. As power adjustments are relatively small ($\sigma_c \approx 1$), generator limits are not explicitly modeled.

\subsection{Flows}
\label{ss:dcpf} With the base case holding the index $c=0$ and having $\pmb{\hat{d}}
_{0}=\pmb{d}$ and $\pmb{\hat{g}}_{0}=\pmb{g}$, the flows in the grid are
calculated for all cases using the classical DC power flow formulation (\ref{eq:dcpf}). An AC formulation is obtained by making the relevant substitutions from DC to AC power flow, but this will naturally prevent its expression as a MILP.

$\forall c \in \mathcal{C}$:
\begin{subequations}
    \begin{align}
        \pmb{f}_{c}\odot \pmb{w}_{c}                                              & = \pmb{0}\label{subeq:dcpf openings}                                  \\
        f_{c,e}= b_{e}\left(\phi_{c,\text{dst}(e)}- \phi_{c,\text{org}(e)}\right) & \odot (1-w_{c,e})                                                      & \forall e \in \mathcal{E}\label{subeq:dcpf flows} \\
        \pmb{A}\pmb{f}_{c}                                                        & = \pmb{\hat{g}}_{c}- \pmb{\hat{d}}_{c}\label{subeq:dcpf conservation} \\
        \left|\pmb{f}_{c}\right|                                                  & \preceq \overline{\pmb{f}}\label{subeq:dcpf limits}
    \end{align}
    \label{eq:dcpf}
\end{subequations}
When a branch is open, its flow shall be null (\ref{subeq:dcpf
openings}). The flow in the branch $e$ results from the phase angle difference between
its buses (\ref{subeq:dcpf flows}) when the branch is closed. (\ref{subeq:dcpf
conservation}) expresses the power balance in the buses (and implies \eqref{subeq:balancing overall}). Finally, (\ref{subeq:dcpf
limits}) limits the flows in the branches to their respective operational limits.

\subsection{Cost function}
\label{ss:cost function} The cost function to be minimized (\ref{eq:cost
function}) consists of the risk due to de-energized nodes. We define the risk of a single contingency $c$ as its probability $\mathfrak{p}_{c}$ of occurrence during the relevant operating window, multiplied
by the volume of the consequent load loss $\langle 1,\pmb{d}- \hat{\pmb{d}}_{c}\rangle$. Other definitions of the risk, for example, also involving the duration or costs of loss of generation, could be considered.
\begin{equation}
    \min_{\pmb{v}}\sum_{c \in \mathcal{C}^\star} \mathfrak{p}_{c} \langle 1, \pmb
    {d}- \hat{\pmb{d}}_{c} \rangle \label{eq:cost function}
\end{equation}
The risk-based OTS problem minimizes the sum of contingency risks for the chosen switch configuration $\pmb{v}$.

\subsection{Big-M formulation}
\label{ss:bigM} Some equations are not pure MILP operations. This subsection gathers
a reformulation of them for an MILP solver using the big-M technique, where $M$ is
a sufficiently large scalar.

The equations involving Hadamard products of variables in the formulation (\ref{subeq:connecteness
branch opening}), (\ref{subeq:N-1 connecteness branch opening}), (\ref{subeq:balancing generation}) and (\ref{subeq:dcpf openings}), generalize
under the form $\pmb{a}= k \pmb{b}\odot \pmb{u}$, with $k$ a scalar and
$\pmb{u}$ an indicator vector and $\pmb{a}$ and $\pmb{b}$ vectors of reals. It translates
as follows:
\begin{subequations}
    \begin{align}
        \pmb{a}            & \preceq M\pmb{u}            \\
        -\pmb{a}           & \preceq M\pmb{u}            \\
        \pmb{a}- k\pmb{b}  & \preceq M(\pmb{1}- \pmb{u}) \\
        -\pmb{a}+ k\pmb{b} & \preceq M(\pmb{1}- \pmb{u})
    \end{align}
\end{subequations}
The absolute value in (\ref{subeq:dcpf limits}) translates in
\begin{subequations}
    \begin{align}
        \pmb{f}_{c} & \preceq \overline{\pmb{f}}  \\
        \pmb{f}_{c} & \preceq -\overline{\pmb{f}}
    \end{align}
\end{subequations}
The last equation to be adapted is the logical expression (\ref{subeq:N-1
connectedness pi disjunction}). To do so, the following translation of the
logical operators on binary variables will be necessary:
\begin{subequations}
    \begin{align}
        x = a \vee b   & \Leftrightarrow \begin{cases}x \ge a \\ x \ge b \\ x \le a+b\end{cases}   \\
        x = a \wedge b & \Leftrightarrow \begin{cases}x \le a \\ x \le b \\ x \ge a+b-1\end{cases}
    \end{align}\label{eq:logical operators to MILP}
\end{subequations}
First, we introduce an intermediate binary variable $\psi_{c,i,e}$ which is
defined
$\forall c \in \mathcal{C}, \forall i \in \mathcal{V}, \forall e \in \text{inc}(i
)$
\begin{equation}
    \psi_{c,i,e}= \pi_{c,\text{opp}(i,e)}\wedge (1-w_{c,e})
\end{equation}
which translates using (\ref{eq:logical operators to MILP}), into
\begin{subequations}
    \allowdisplaybreaks
    \begin{align}
        \psi_{c,i,e} & \le \pi_{c,\text{opp}(i,e)}            \\
        \psi_{c,i,e} & \le 1 - w_{c,e}\label{eq:BigM Mix-off} \\
        \psi_{c,i,e} & \ge \pi_{c,\text{opp}(i,e)}- w_{c,e}
    \end{align}
\end{subequations}
Now (\ref{subeq:N-1 connectedness pi disjunction}) becomes
\begin{equation}
    \pi_{c,i}= \bigvee_{e \in \text{inc}(i)}\psi_{c,i,e}\quad \forall (c,i) \in \mathcal{C}
    \times\mathcal{V}
\end{equation}

which translates using (\ref{eq:logical operators to MILP}) into

$\forall (c,i) \in \mathcal{C}\times\mathcal{V}$:
\begin{subequations}
    \allowdisplaybreaks
    \begin{align}
        \pi_{c,i} & \le \sum_{e \in \text{inc}(i) }\psi_{c,i,e} \\
        \pi_{c,i} & \ge \psi_{c,i,e}                           & \forall e \in \text{inc}(i)
    \end{align}
\end{subequations}

%% file: results.tex
\section{Results}
\begin{figure}[h]
    \begin{subfigure}
        [t]{.43\linewidth}
        \includegraphics[width=\linewidth]{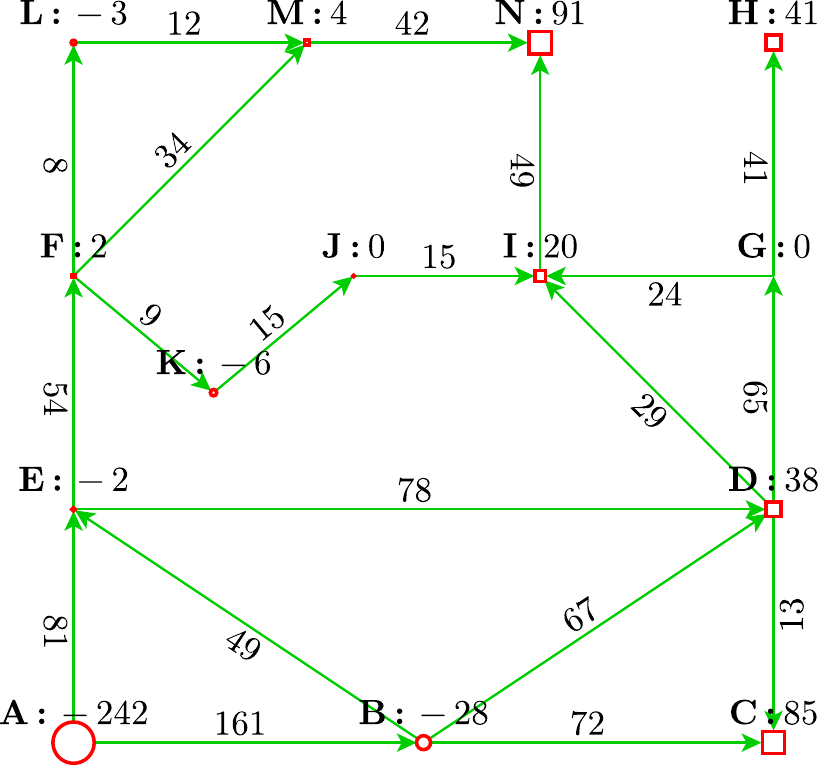}
        \caption{base case}
        \label{fig:case base}
    \end{subfigure}\hfill
    \begin{subfigure}
        [t]{.48\linewidth}
        \includegraphics[width=\linewidth]{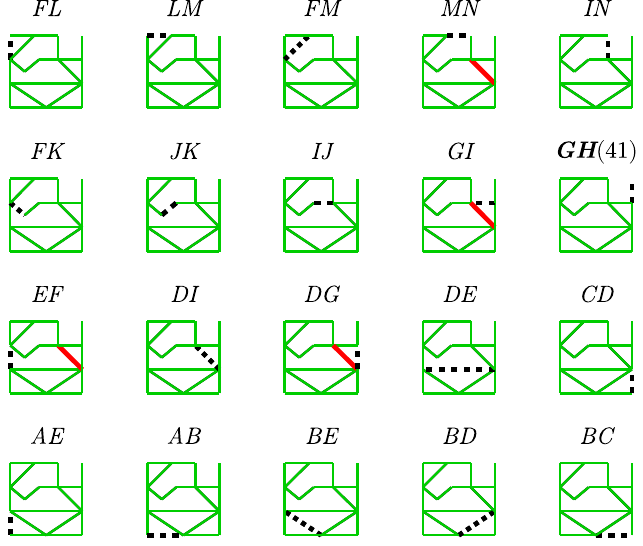}
        \caption{security analysis}
        \label{fig:grid base}
    \end{subfigure}
    \caption{Base without OTS: There is no overflow in the base case. The
    security analysis reveals that numerous trippings raise an overloading on $ID$.}
    \label{fig:base}
\end{figure}

The model was applied to the IEEE 14-bus network and a single loading scenario, shown in Fig.~\ref{fig:base}. The contingency list contained all
branches, that is, $\mathcal{C}= \mathcal{E}$, and for this initial study all contingencies were assumed to be equally likely. It was modeled in Julia and solved by Gurobi 11 on an Apple M3. The computation time was 1.8 seconds, which in Gurobi's metric corresponds to 3.26 work units.

\begin{figure}[h]
    \begin{subfigure}
        [t]{.43\linewidth}
        \includegraphics[width=\linewidth]{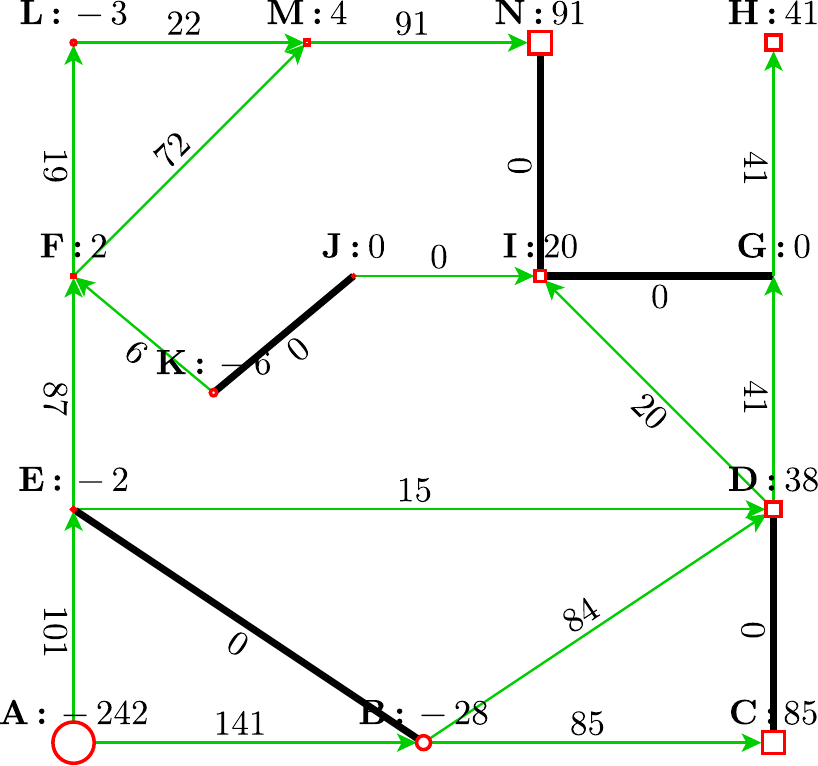}
        \caption{base case}
        \label{fig:case ots}
    \end{subfigure}\hfill
    \begin{subfigure}
        [t]{.48\linewidth}
        \includegraphics[width=\linewidth]{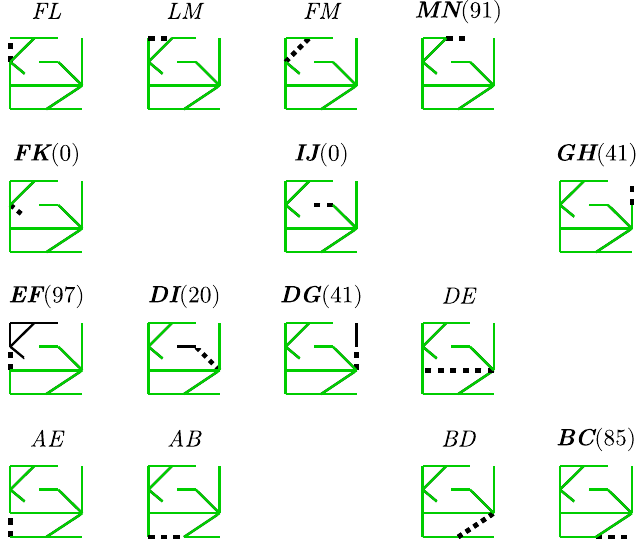}
        \caption{security analysis}
        \label{fig:grid ots}
    \end{subfigure}
    \caption{Base case with OTS.}
    \label{fig:ots}
\end{figure}

\begin{figure}[h]
    \begin{subfigure}
        [t]{.43\linewidth}
        \includegraphics[width=\linewidth]{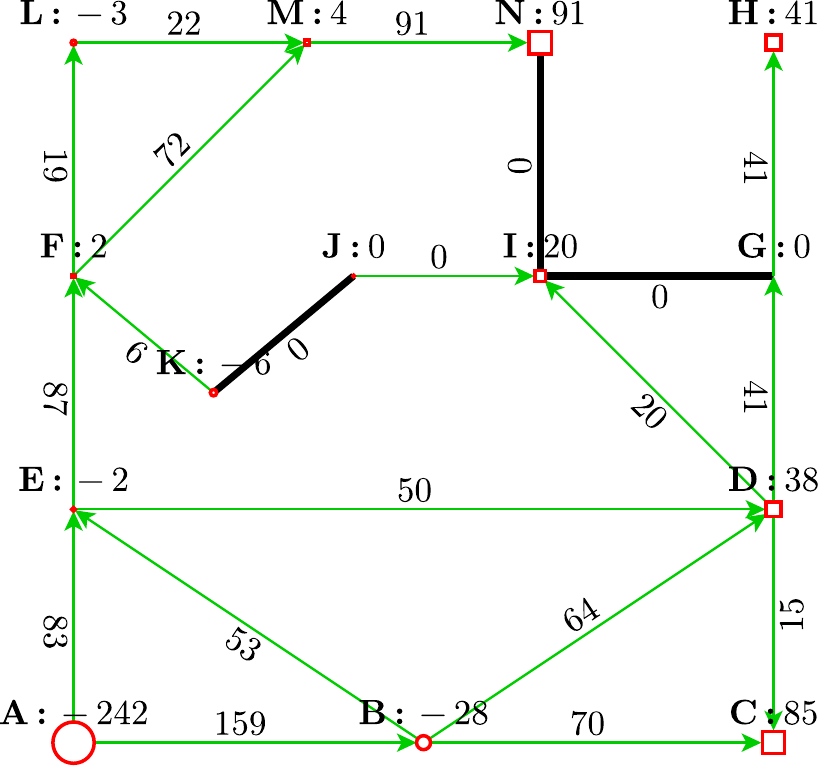}
        \caption{base case}
        \label{fig:case intermediate}
    \end{subfigure}\hfill
    \begin{subfigure}
        [t]{.48\linewidth}
        \includegraphics[width=\linewidth]{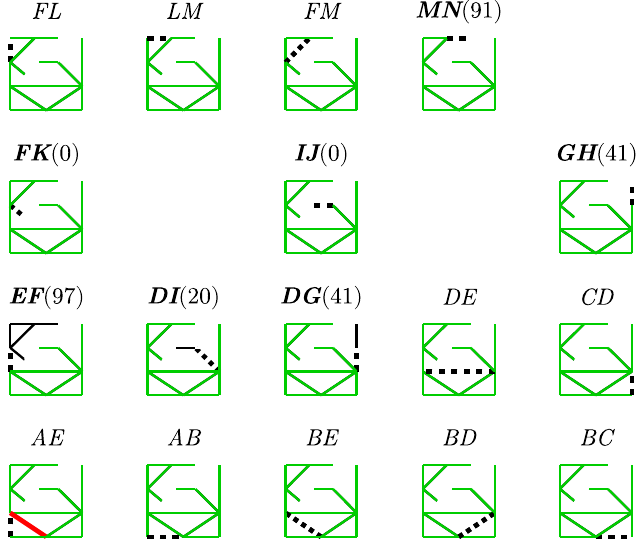}
        \caption{security analysis}
        \label{fig:grid intermediate}
    \end{subfigure}
    \caption{Intermediate case the operator would needed to step into. The branch
    openings in the northern area solve the constraint on branch $DI$, but
    raises one on branch $BE$. }
    \label{fig:intermediate}
\end{figure}

Figs.~\ref{fig:base}-\ref{fig:intermediate} each consist of two panels. The pre-contingency scenario is shown on the left. 
Buses represented as circles are
generating, while those represented as squares are consuming. Their sizes follow
the absolute value of their injections, and net consumption (in MW) is given. Branches are color coded as black (opened by the OTS algorithm) or green (closed), where arrows and numbers indicate the flow (in MW). 

The security analysis is shown on the right of each figure. The title of each subfigure
is the name of the branch that tripped. It is in bold style when a part of the
grid is de-energized and the load lost is in brackets. 
In the security analysis figures, the lines opened
by the OTS algorithm are removed, dashed black lines are those the contingency one, and solid black lines those that are in a de-energized area. The already opened lines are removed and overloaded lines are shown in red.

The base case is shown in Figure \ref{fig:case base}. There is no overloading; therefore,
if branches are to be open, it is only to cope with overloads in N-1. The source
vertex corresponds to the bus $A$, which holds the largest generator. Figure \ref{fig:grid
base} shows the results of the security analysis where 4 of 20 trippings end with
the overloading of the branch $DI$. As $GH$ forms an antenna, its tripping results in
the loss of the bus $H$.

\begin{table}[htbp]
\caption{Result summary}
\begin{center}
\begin{tabular}{|c|c|c|c|}
\hline
\textbf{}&\multicolumn{3}{|c|}{\textbf{Scenario}} \\
\cline{2-4} 
\textbf{metric} & \textbf{\textit{base}}& \textbf{\textit{intermediate}}& \textbf{\textit{optimized}} \\
\hline
preventive branch openings & 0 & 3 & 5 \\
    contingencies with overloads & 4 & 1 & 0\\
     contingencies with de-energized buses & n/a$^{\mathrm{a}}$ & n/a$^{\mathrm{a}}$ & 8 \\
     average demand loss, per contingency & n/a$^{\mathrm{a}}$ & n/a$^{\mathrm{a}}$ & 6.7\% \\
\hline
\multicolumn{4}{l}{$^{\mathrm{a}}$due to overloads, consequences are unknown.}
\end{tabular}
\label{tab1}
\end{center}
\end{table}

Figure \ref{fig:case ots} shows the situation with the branch opening proposed
by our algorithm. As expected, there is no overloading anymore as shown by the
security analysis in Figure \ref{fig:grid ots}, but this comes at the cost of loss of
buses for for 8 of 14 contingencies. Now, $DI$ only holds an antenna with the load
of the buses $I$ and $J$ and its load will never exceed the sum of both consumptions.
However, to isolate this antenna, $IG$ is opened, which grows the previously existing
antenna with $G$ and $H$. In addition, $IN$ and $IJ$ are opened, creating a
peninsula held on branch $EF$ with the buses $F$, $K$, $L$, $M$ and $N$.

That scenario also reveals the preventive-openings-cascade phenomenon. In fact, in the
base case, only the branch $DI$ is at risk of being overloaded and the opening of
$JK$, $IN$ and $GI$ may appear at first sufficient. Figure
\ref{fig:intermediate} shows the intermediate state if the operator only focuses
on the initial constraint. None of the trippings in the northern area leads to overloads.
But that reconfiguration also affected the distribution of the flows in the
southern part, which raises a new constraint. Now, the tripping of $AE$, which
was previously sound, ends with the overloading of the branch $BE$. Thus, resolving
the constraint in the northern area weakens the southern one, and consequently a
reconfiguration becomes also necessary in that area: the opening of the branches
$BE$, and $CD$, and now $BC$ becomes an antenna.

This rationale is similar to what would have been necessary for an operator facing
the same situation. Sometimes back and forth steps are necessary. This kind of iterative
approach is time consuming even for such a simple case, and the operator would
settle for a solution that respects the security rules, even if it may be far from
optimal. Thanks to the optimization technique, all the constraints are incorporated
into one single problem and then solved at once. Moreover, by taking into account
the probability associated with the tripping of each individual line, as well as
the subsequent loss, a solution can be found that completely avoids the exposure to the unknown risks associated with overloading and minimizes the risk related to de-energized buses. 
The performance is summarized in Table~\ref{tab1}.

%% file: conclusion.tex
\section{Conclusion}
Noting that the N-1 rule is fundamentally a risk-based approach, our approach highlights situations that necessarily happen in transmission grids, and especially in subtransmission systems where the risk of de-energizing parts of the grid following a single contingency is considered. This operational policy is modeled as an MILP, which leverages connectivity analysis in the N-1 cases that is incorporated into the OTS problem. As a result, solutions similar to those observed in operation are reached by the algorithm, notably revealing the preventive-openings-cascade phenomenon. The application of such an approach would benefit the operator, as the design of these strategies is time consuming. 

In future work, this approach will be extended with bus splitting actions that are generally preferred to line openings, AC power flow to take into account reactive power and voltage limitations, and completed with an Optimal Power Flow to include the redispatching lever. Finally, the algorithm must scale for solving the problem on real-size grids.